\documentstyle[12pt]{article}
\begin{document}
\addtolength\leftmargin{-2cm}
\title{\bf \large CONFINEMENT AND THE TRANSVERSE LATTICE\footnote{
Invited talk, presented at the 1995 ELFE meeting on confinement physics,
Cambridge, UK, July 1995.}}
\author{\normalsize Matthias Burkardt\\[-.5ex]
\normalsize Physics Dept., New Mexico State University, Las Cruces, NM
88003-0001}
\date{}
\thispagestyle{empty}
\maketitle
\begin{small}\noindent
The status of the transverse lattice formulation of light-front QCD is
reviewed.
It is explained how confinement arises in this formulation for large lattice
spacing. The
nonperturbative re\-nor\-ma\-li\-za\-tion procedure is outlined in general and
a
more
detailed discussion is given for the case of $QCD_{2+1}(N_C\rightarrow
\infty)$.
\end{small}
\section{Introduction}
Deep inelastic lepton-hadron scattering has played a
fundamental role in the investigation of hadron
structure. For example, the discovery of Bjorken scaling
confirmed the existence of point-like charged
objects inside the nucleon (quarks).
Besides such fundamental discoveries, deep inelastic
scattering (DIS)
revealed surprising and interesting details
about the structure of nucleons and nuclei, such as
\begin{itemize}
\item the nuclear EMC effect
\item the spin crisis in polarized DIS experiments
(EMC, NMC, SMC, SLAC, E665)
\item the isospin asymmetry of the nucleon's Dirac sea (NMC)\\
$\hookrightarrow$ failure of the Gottfried sum-rule
\end{itemize}

More interesting data is expected in the near future:
For example, the experiments by the HERMES collaboration
at HERA will provide
measurements of polarized structure functions (leading and higher twist)
with unprecedented
precision. Chirally odd parton distributions will be accessible
both at HERMES and in proton-proton collisions at the
RHIC facility. Future experiments at CEBAF-II and the
planned European high energy accelerator ELFE
will provide additional information about semi-inclusive
DIS and higher-twist effects.

Perturbative QCD evolution has been successfully applied to
correlate large amounts of experimental data.
However, progress in understanding nonperturbative
features of the parton distributions in these experiments
has been slow. In fact, the theoretical understanding of
the surprising results listed above is mostly limited
to {\it ad hoc} models with little connection to the
underlying quark and gluon degrees of freedom.

Part of the difficulty in describing parton distributions
nonperturbatively derives from the fact that parton
distributions measured in DIS are dominated by
correlations along the light-cone ($x^2=0$).
For example, this makes calculations of parton distributions on
a Euclidean lattice, where all distances are
space-like, exceedingly difficult.
Furthermore, in an equal time quantization scheme,
deep inelastic structure functions are described by
real time response functions which are not only
very difficult to interpret but also to calculate.

Light-Front (LF) quantization seems is a promising
tool to describe the immense wealth of experimental
information about structure functions for a variety
of reasons:
\begin{itemize}
\item correlations along the light-cone become ``static'' observables
in this approach
[i.e. equal $x^+ \equiv (x^0+x^3)/\sqrt{2}$ observables]
\item structure functions are easy to evaluate from the
LF wavefunctions
\item structure functions are easily interpreted as LF
momentum densities
\end{itemize}
Further advantages of the LF formalism derive from the
simplified vacuum structure (nontrivial vacuum effects
can only appear in zero-mode degrees of freedom) which
provides a physical basis for the description of
hadrons that stays close to intuition.

\section{The Transverse Lattice}
\begin{figure}
\begin{Large}
\unitlength.8cm
\begin{picture}(14,9.3)(-2,5.5)
\put(1.5,8.5){\line(0,1){1.7}}
\put(1.,11.1){\makebox(0,0){(discrete)}}
\put(1.,12.0){\makebox(0,0){$\perp$ space}}
\put(1.5,12.6){\vector(0,1){1.6}}
\put(2.,8.){\line(3,-1){1.2}}
\put(4.5,6.9){\makebox(0,0){long. space}}
\put(4.5,6.2){\makebox(0,0){(continuous)}}
\put(7.1,6.3){\vector(3,-1){1.8}}
\put(10.4,5.8){\line(3,1){1.8}}
\put(14.5,6.2){\makebox(0,0){(continuous)}}
\put(14.5,6.9){\makebox(0,0){time}}
\put(15.8,7.6){\vector(3,1){1.8}}
\includegraphics{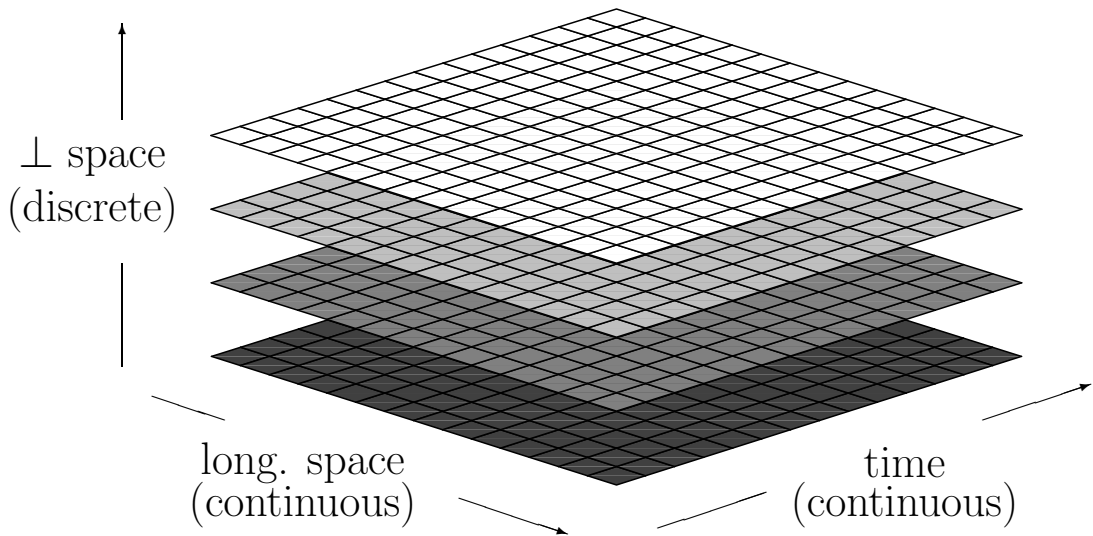}
\end{picture}
\end{Large}
\caption{Space time view of a transverse lattice}
\label{fig:perpl}
\end{figure}

Before one can apply the LF formalism to QCD one has to remove
the divergences first (i.e. regularize and renormalize).
Then one has to formulate bound state problems into a
form that can be solved numerically with a reasonable
effort.
The basic idea of the transverse lattice  --- an idea originally suggested
by Bardeen, Pearson and Rabinovici \cite{bpr} ---is very simple:
one keeps two directions (the time and the z-direction)
continuous but discretizes the transverse space coordinates
(Fig. \ref{fig:perpl}). The metric is Minkowskian.
Two immediate advantages of this construction
are
\begin{itemize}
\item manifest boost and translational invariance
in the longitudinal direction --- thus keeping
parton distributions easily accessible
\item a gauge invariant cutoff for divergences
associated with large transverse momenta
\end{itemize}
Several steps are required before one can use this
construction to actually calculate parton distribution
functions:

First, one has to write down a discretized approximation
to the QCD action on a $\perp$ lattice. This has been done
in the original work \cite{bpr}. Since space is kept
continuous in the $\perp$ direction, the longitudinal
component of the gauge field is non-compact. However, in
order to maintain gauge invariance in the $\perp$ direction,
one introduces compact $SU(N_C)$ fields to represent the
$\perp$ component of the gluon field.
One can thus think
of the $\perp$ lattice as a large number \footnote{The number
is equal to the number of transverse links.} of gauged,
nonlinear $\sigma$-model fields in 1+1 dimensions
coupled together. This picture will prove useful
when one performs nonperturbative numerical calculations.
Naive fermions are straightforward to implement:
the only difference between the continuum action and the
$\perp$ lattice action is the replacement of
$\perp$ derivatives by finite differences with appropriate
gauge field links to maintain gauge invariance.

Note that, since species doubling occurs only for the
two latticized transverse directions, species
doubling is a lesser problem than on a Euclidean lattice
and only 4 species of fermions arise if one employs
a naive fermion action. This can be easily dealt
with, using staggered fermions \cite{double}.
An alternative to staggering the fermion degrees
of freedom is introducing additional terms in the
action that lift the degeneracy between the naive
species (Wilson-formulation). There are some subtle
differences between chiral symmetry on the LF and what one usually
considers chiral symmetry (the difference arises from
the fact that on the LF only half the degrees of freedom are
treated as dynamical degrees of freedom) \cite{osu}.
It turns out that Wilson fermions on the LF do not
break the LF-chiral symmetry \cite{wilson}, which
makes them very appealing for practical calculations.
However, I should emphasize that this does {\it not}
solve the general problem of formulating chiral
theories on a lattice since ``chiral'' transformations
in the two component LF-formulation are slightly different
from  usual chiral transformations \cite{osu}.

In the next step, one applies LF quantization to the
transverse lattice action. Once one has thus derived
a LF Hamiltonian, one can proceed to the final step
and apply powerful numerical techniques to solve
for the low lying eigenstates (= low lying hadrons)
of the LF-Hamiltonian. From the ground state
(LF-) wavefunctions one can then easily extract
parton distribution functions and other
light-like correlation functions (both leading and higher twist!).

It is interesting to see how confinement emerges on the transverse lattice
in the limit of large lattice spacing:
In this limit, the coupling between the sheets is weak and the
energy scale associated with link field excitations is high.
Thus, when one separates two test charges in the longitudinal direction,
the transverse lattice behaves similar to QCD$_{1+1}$ and linear
confinement results trivially. For transverse separations between the
charges, a different mechanism is at work. Gauge invariance demands
that the two charges are connected by a string of link fields.
For large spacing the link fields fluctuate only little
and the energy of such a configuration can be estimated by counting the
number of link fields needed to connect the charges, which again yields
linear confinement \cite{review}.
Of course, as the spacing gets smaller, confinement becomes a
dynamical question. However, it is a tremendous practical
advantage if there is a limit which is technically simple but
still very close to the expected physics.

\section{Monte Carlo Techniques}
In a recent paper \cite{lfepmc}, I have demonstrated that
transverse lattice Hamiltonians can be solved using
powerful Monte-Carlo techniques. The basic idea is very simple:
since the LF Hamiltonian is local in the $\perp$ direction
(only nearest neighbor interactions) one can set up
the problem of finding ground states in such a way that
one first solves the Hamiltonian numerically exactly
on one link or one plaquette (this is relatively easy, since it
implies only solving a 1+1 dimensional field theory\footnote{Several
methods to solve 1+1 dimensional field theories in LF quantization
have been described in the literature. In the context of the $\perp$
lattice, DLCQ \cite{peb} seems to be quite useful.}
).
Then one includes the coupling in the transverse direction
by means of a {\it random walk in Fock space} governed
by ensemble projector Monte Carlo techniques.
Such algorithms eliminate the need for
Tamm-Dancoff approximations and thus
allow to solve reasonably large $\perp$ lattices on
realistic computers. For example, in Ref.\cite{lfepmc},
I solved the $\perp$ lattice LF Hamiltonian for
self-interacting scalars on 64 $\perp$ sites on a
workstation. Even if one accounts for the increased
complexity of the Hamiltonian in a gauge theory,
it thus seems feasible to perform QCD calculations on a
similar size $\perp$ lattice on available supercomputers.

\section{Renormalization}
Due to the lack of full covariance (specifically rotational invariance),
renormalization
on a $\perp$ lattice is anything but trivial. Since the
cutoffs violate rotational invariance one must also allow
for counter terms which (by themselves) violate rotational
invariance in order to achieve rotational invariance in physical
observables \cite{mb:rot}. The richer counter term
structure is compensated by an increase in renormalization
conditions. The infinite parts of various counter terms are often
related when a symmetry is {\it hidden}. This phenomenon
has been called {\it coupling constant coherence}
and has been used in the context of light-front Tamm Dancoff
\cite{ccc}. It should also be of use in the context
of the transverse lattice.
The finite part of the symmetry breaking (or restoring!) counter terms
can be determined by making use of the increased number of renormalization
conditions.
At least in principle, one can demand
full covariance for each physical observable. This
provides a very large number (in principle infinite)
of renormalization conditions which one can use to
determine the (finite parts of) non-covariant counter terms.
An observable which is particularly useful in this
context is the potential between infinitely heavy
quarks. In Ref.\cite{mb:zako}, I have described how
one can set up a numerical calculation in the context
of LF quantization which calculates the $Q\bar{Q}$ potential.
Currently, I am performing numerical test calculations
on a 2+1 dimensional lattice for $QCD(N_C\rightarrow \infty)$
using a Lanczos algorithm. In these calculations I am
fine-tuning the finite part of the counter terms such that
one obtains a rotationally invariant $Q\bar{Q}$ potential.

Vacuum effects, i.e. effects that are associated with
zero-modes in the LF formulation, were largely ignored
in Ref.\cite{bpr}. There has been substantial progress
in understanding how zero modes affect LF Hamiltonians
in the continuum limit (see e.g. Refs.\cite{review,mb:sg,ro93,kp93,world} and
references therein). The general result of these studies is
that the LF vacuum in the continuum limit is trivial but that
one has to pay for this advantage with a more complicated
Hamiltonian. However, on the $\perp$ lattice only a few additional
terms are necessary.

In addition to these renormalization issues that are typical for
any LF field theory, one must also address issues that are specific
for the transverse lattice formulation, such as {\it scale setting}.
This will be illustrated for the example of $QCD_{2+1}(N_C\rightarrow \infty)$
\footnote{Glueball spectra for $QCD_{2+1}(N_C\rightarrow \infty)$ in the
Euclidean formulation are available by taking numerical results from
$N_C=2,3,4$
\cite{teper} and extrapolating them to $N_C\rightarrow \infty$},
where the LF Hamiltonian reads\cite{bpr}
\begin{eqnarray}
P^- =\sum_n & &\left[-c_g\int dx^- dy^-
J^{ij}_{n,n+1}(x^-)|x^--y^-|J^{ji}_{n,n+1}(y^-)\right.
\\
& &+\left.c_4 \int dx^- tr \left(U_n(x^-)U^\dagger_n(x^-)
U_n(x^-)U^\dagger_n(x^-)\right)
+c_2 \int dx^-tr\left(U_n(x^-)U^\dagger_n(x^-)\right)\right],
\nonumber
\end{eqnarray}
where $U_n(x^-)\equiv U^{ij}_n(x^-)$ is the link field, represented by a
complex
$N_C \times N_C$ matrix,
\begin{equation}
J^{ij}_{n,n+1}(x^-)=
U_n^\dagger(x^-) \stackrel{\leftrightarrow}{\partial_-} U_n(x^-) -
U_{n+1}(x^-) \stackrel{\leftrightarrow}{\partial_-}
U_{n+1}^\dagger(x^-)
\end{equation}
is the current that couples to the longitudinal gauge filed at each site
and the terms multiplied by the constants $c_4$ and $c_2$ are introduced to
enforce the $SU(N_C)$ constraint on the link fields $U^{ij}_n$.
$c_g$ is the gauge coupling within the $1+1$ dimensional sheets.
In the naive continuum limit, it is related to the $2+1$ dimensional
gauge coupling via $c_g=g^2/a$, i.e. it carries  dimension
$L^{-2}$.

The coupling constants are chosen as follows:
First of all, $c_4$ and $c_2$ are supposed to enforce
the $SU(N_C)$ constraint, which implies in the classical
action $c_2=-2c_4$ and $c_4\rightarrow \infty$.
Due to renormalization effects (tadpoles etc.\cite{mb:sg})
$c_2$ picks up additional pieces so that this procedure has to be
modified a little: one still takes (gradually)
$c_4\rightarrow \infty$.
For every value of $c_4$, $c_2$ is chosen near the
critical point, where the transverse string tension\footnote{See
\cite{mb:zako} for an algorithm to calculate
longitudinal and transverse string tensions from
a LF Hamiltonian.}
vanishes in lattice units: For large values of
$c_2$, the energy of a transversely separated
static quark anti-quark pair grows roughly linearly
with the number of sites in between the $Q\bar{Q}$
pair. As one decreases $c_2$, the growth with the
number of sites becomes slower and slower, which means
that the lattice spacing, in units of the physical
string tension, decreases. The continuum limit, i.e.
vanishing lattice spacing $a$, has been reached when the
energy of the $Q\bar{Q}$ pair no longer increases when
their separation increases in lattice units, which
is identical with the critical point beyond which the
LF spectrum becomes tachionic.
In practice, this means that if one is not interested in
the actual numerical value of the lattice spacing in
physical units, then one does not need to evaluate the
transverse string tension! One only has to make sure that one
is close to the critical point. Of course, the transverse
string tension is nevertheless interesting because one
would like to understand how confinement arises on the
transverse lattice in the continuum limit.

So far, we have not discussed how one fixes the longitudinal
gauge coupling $c_g$. It plays a dual role: first, at least
if one measures the other couplings in units of $c_g$, it merely fixes the
overall mass scale, and
its numerical value is irrelevant as long as one always
forms dimensionless combinations of physical observables,
such as ratios of masses and/or the string tension.
The second role played by $c_g$ is that it largely determines the
longitudinal string tension.
For a $Q\bar{Q}$ pair separated
on the same site in the longitudinal direction no gluons are
required to maintain residual gauge
invariance. Since the valence approximation
is usually rather good in LF calculations this implies that
the longitudinal string tension roughly equals  $c_g$.
In summary, this implies the following nonperturbative renormalization
procedure:
\begin{itemize}
\item[1] pick an arbitrary value for $c_g$
\item[2] pick an arbitrary value for $c_4$
\item[3] solve the spectrum as a function of $c_2$ and tune $c_2$ such that
$c_2$ stays slightly above the critical point.
\item[4] increase $c_4$ and continue with step 3 until one is close
enough to the continuum limit. There are at least two ways to verify whether or
not one is close to the continuum.
One is to actually calculate the string tension in lattice units. The other is
to study numerical convergence of dimensionless quantities, such as mass
ratios.
\end{itemize}
In practice, a different approach is also conceivable: instead of trying
to make the lattice spacing as small as possibly (i.e. working very
close to the critical line), it may be more efficient to look for the
{\it perfect LF-Hamiltonian} for finite spacing (i.e. coupling constants
away from the critical line). This implies that one has to add more terms
to the LF Hamiltonian and the nonperturbative renormalization procedure
becomes more complex, which is certainly a disadvantage.
However, the advantage of such a procedure would be that one stays
away from the critical regime. Since the Fock space content of hadrons
is known to become infinitely complicated very close to the critical
regime, this implies that a perfect LF-Hamiltonian can be more
easily solved numerically and in the end, be more economical.
\section{Summary}
The transverse lattice is a hybrid technique (a shotgun marriage between
lattice
field theory and light-front field theory)
to solve relativistic field theories.
Its main advantages over conventional lattice gauge theories is that
parton distributions are much more easily calculable and interpretable. The
disadvantage compared to conventional lattice
gauge theories is that one inherits some (not all) problems
of light-front field theories. The advantages over conventional
light-front field theory is manifest invariance under a class of
residual gauge transformation, less infrared problems
and simple ways to implement Monte Carlo algorithms.
Furthermore, confinement is ``built in'', which means that
the zeroth order approximation is already close to the expected
physics in QCD.
The disadvantages compared to other approaches to
light-front field theory (say LF Tamm Dancoff) is an
increase in the complexity of the wave functions.
Current results on the transverse lattice are still limited
to simple model studies in $2+1$ dimensions, but the reason is
more lack of manpower than any physics reason.
Considering the substantial experimental effort to explore
the parton structure of hadrons, it is definitely worthwhile
to investigate further, to what extend one can describe
deep inelastic structure functions using the transverse lattice
formulation of QCD.

\end{document}